\documentclass[11pt]{article}
\usepackage[textwidth=15.2cm,textheight=22cm]{geometry}
\usepackage{amsmath,amssymb}
\usepackage{latexsym}
\usepackage{multicol}
\usepackage{graphicx}
\usepackage{bm}
\allowdisplaybreaks[1]
\numberwithin{equation}{section}

\newcommand{\del}{\partial}
\newcommand{\be}{\begin{equation}}
\newcommand{\ee}{\end{equation}}
\newcommand{\ba}{\begin{eqnarray}}
\newcommand{\ea}{\end{eqnarray}}
\newcommand{\bdm}{\begin{displaymath}}

\newcommand{\edm}{\end{displaymath}}

\newcommand{\rom}[1]{\uppercase\expandafter{\romannumeral #1\relax}}
\renewcommand{\d}{\partial}

\newcommand{\E}{E_{7(7)}}
\def\parp{\partial^+}

\def\d{\partial}

\def\ba{\bar A}

\def\m#1{\mathcal#1}
\def\beq{\begin{equation}}
\def\eeq{\end{equation}}

\newcommand{\nn}{\nonumber}

\newcommand{\ndt}{\noindent}

\newcommand{\delp}{{\partial^+}}

\def\bea{\begin{eqnarray}}
\def\eea{\end{eqnarray}}
\def\beas{\begin{eqnarray*}}
\def\eeas{\end{eqnarray*}}
\def\sla{\raise.15ex\hbox{$/$}\kern-.57em}

\def\parm{{\partial}_{-}}
\def\parp{\partial^+}

\def\spa#1.#2{\left\langle#1\,#2\right\rangle}
\def\spb#1.#2{\left[#1\,#2\right]}

\begin{document}

\begin{titlepage}
\begin{flushright}    
{\small $\,$}
\end{flushright}
\vskip 1cm
\centerline{\Large{\bf{$E_8$ in $\mathcal N=8$ supergravity in four dimensions}}}
\vskip 1.5cm
\centerline{Sudarshan Ananth$^\dagger$, Lars Brink$^*$ and Sucheta Majumdar$^\dagger$}
\vskip .7cm
\centerline{$\dagger$\,\it {Indian Institute of Science Education and Research}}
\centerline{\it {Pune 411008, India}}
\vskip 0.7cm
\centerline{$^*\,$\it {Department of  Physics, Chalmers University of Technology}}
\centerline{\it {S-41296 G\"oteborg, Sweden}}
\vskip 0.1cm
\centerline{\it {and}}
\vskip 0.1cm
\centerline{\it{Division of Physics and Applied Physics, School of Physical and Mathematical Sciences}}
\centerline{\it{Nanyang Technological University, Singapore 637371}}
\vskip 1.5cm
\centerline{\bf {Abstract}}
\vskip .5cm

\ndt We argue that $\mathcal N=8$ supergravity in four dimensions exhibits an exceptional  $E_{8(8)}$ symmetry, enhanced from the known $ E_{7(7)}$ invariance. Our procedure to demonstrate this involves dimensional reduction of the $\mathcal N=8$ theory to $d=3$, a field redefinition to render the $E_{8(8)}$ invariance manifest, followed by dimensional oxidation back to $d=4$. 

\vfill
\end{titlepage}

\section{Introduction}

\vskip 0.5cm
Gravity with maximal supersymmetry in four dimensions, $\mathcal N=8$ supergravity,  exhibits both $\mathcal N=8$ supersymmetry and the exceptional  $ E_{7(7)}$  symmetry~\cite{Cremmer:1979up}. This theory is known to be better behaved in the ultraviolet than pure gravity and has recently been shown to be finite up to four loops~\cite{Bern:2011qn}. There is mounting evidence from these calculations and others that points to unexpected cancellations and hence an underlying enhanced symmetry. Using light-cone superspace, the Hamiltonian of maximal supergravity, in $d=4$, is constructed as a power series in the coupling constant and this has been achieved up to the four-point coupling. In an earlier paper~\cite{Ananth:2016abv}, we showed that ``oxidation"~\cite{oxid} of the $\mathcal N=8$ theory to $d=11$, suggests that there is an $ E_{7(7)}$ symmetry in eleven dimensions. This result has been shown to first order in the coupling constant. Since the the states of the $d=11$ theory are not representations of the linearly realized maximal subgroup $SU(8)$ of the  $ E_{7(7)}$ symmetry they have to be broken up into such representations to see the symmetry. This is accomplished by using the same superfield in all dimensions (note that the number of states is always $128$ bosons $+$ $128$ fermions.) By writing the Hamiltonian in this formulation we can prove the symmetry to the lowest order in the gravitational coupling constant.
\vskip 0.3cm
\ndt Motivated by this, we describe in this paper an entire process involving dimensional reduction, field redefinitions and dimensional oxidation that leads us to conclude that $\mathcal N=8$ supergravity in $d=4$ exhibits an exceptional $E_{8(8)}$ symmetry, at least to second order in the coupling constant, enhanced from $ E_{7(7)}$.
\vskip 0.3cm
\ndt In section 2, we review the formulation of $(\mathcal N=8, d=4)$ supergravity in light-cone superspace. We dimensionally reduce this $d=4$ theory in a straightforward way to arrive at an action, in three dimensions, which mimics the four-dimensional one but with only one transverse derivative. This formulation thus inherits a three-point coupling and cannot exhibit the maximal subgroup of $ E_{8(8)}$, $SO(16)$, in a linear fashion since under this symmetry, the states of the theory transform as $128$-dimensional spinors. Three such states cannot form a scalar. In the following section, we present a field redefinition that maps this three-dimensional theory, with a cubic vertex to a three-dimensional theory without one. This form of the three-dimensional theory exhibits both $SO(16)$ invariance and a full $E_{8(8)}$ symmetry~\cite{Marcus:1983hb}. We can then go back to the first formulation and indeed find the $SO(16)$ symmetry there, now realized non-linearly. We then ``oxidize" the second formulation back to four dimensions in a manner that preserves all the symmetries discussed earlier, thus arriving at a four-dimensional maximally supersymmetric theory with $E_{8(8)}$ invariance to that order.
\vskip 0.3cm
\ndt Our formulation uses only the real degrees of freedom of the theory. This means that we lose a lot of the covariance usually found in gravity theories, since many of the symmetries are non-linearly realized. In a sense the formulation is packed with symmetries, some of which are difficult to see. By making various field redefinitions we can make particular symmetries visible but one formulation will never be enough to find all the symmetries. We strongly believe that maximal supergravity and Yang-Mills theories have as many symmetries as one can pack into one theory, and this is why they have unique quantum properties.

\section{($\mathcal N=8,d=4$) Supergravity in light-cone superspace}

\vskip 0.2cm

\noindent With the metric $(-,+,+,+)$, the light-cone coordinates are
\bea
{x^{\pm}}=\frac{1}{\sqrt 2}\,(\,{x^0}\,{\pm}\,{x^3}\,) \quad x =\frac{1}{\sqrt 2}\,(\,x^1\,+\,i\,x^2\,) \quad {\bar x}=(x)^*\ ,
\eea
with the corresponding derivatives being $\partial^\mp, \bar\partial$ and $\partial$. The $\mathcal N=8$ superspace is spanned by the Grassmann variables $\theta^m$ and $\bar \theta_m$ ($m=1\,\ldots\,8$ ), the {\bf 8} and $\bar {\bf 8}$ of $SU(8)$ respectively. All 256 physical degrees of freedom in the  theory are captured by the superfield~\cite{Brink:1982pd} 
\bea\label{superfield}
\begin{split}
\phi\,(\,y\,)\,=&\,\frac{1}{{\parp}^2}\,h\,(y)\,+\,i\,\theta^m\,\frac{1}{{\parp}^2}\,{\bar \psi}_m\,(y)\,+\,\frac{i}{2}\,\theta^m\,\theta^n\,\frac{1}{\parp}\,{\bar A}_{mn}\,(y)\ , \\
\;&-\,\frac{1}{3!}\,\theta^m\,\theta^n\,\theta^p\,\frac{1}{\parp}\,{\bar \chi}_{mnp}\,(y)\,-\,\frac{1}{4!}\,\theta^m\,\theta^n\,\theta^p\,\theta^q\,{\bar C}_{mnpq}\,(y)\ , \\
\;&+\,\frac{i}{5!}\,\theta^m\,\theta^n\,\theta^p\,\theta^q\,\theta^r\,\epsilon_{mnpqrstu}\,\chi^{stu}\,(y)\ ,\\
\;&+\,\frac{i}{6!}\,\theta^m\,\theta^n\,\theta^p\,\theta^q\,\theta^r\,\theta^s\,\epsilon_{mnpqrstu}\,\parp\,A^{tu}\,(y)\ ,\\
\,&+\,\frac{1}{7!}\,\theta^m\,\theta^n\,\theta^p\,\theta^q\,\theta^r\,\theta^s\,\theta^t\,\epsilon_{mnpqrstu}\,\parp\,\psi^u\,(y)\ ,\\
\,&+\,\frac{4}{8!}\,\theta^m\,\theta^n\,\theta^p\,\theta^q\,\theta^r\,\theta^s\,\theta^t\,\theta^u\,\epsilon_{mnpqrstu}\,{\parp}^2\,{\bar h}\,(y)\ ,
\end{split}
\eea
\noindent where $h$ and $\bar h$ represent the graviton, ${\bar \psi}_m$ the $8$ spin-$\frac{3}{2}$ gravitinos, ${\bar A}_{mn}$ the $28$ gauge fields, ${\bar \chi}_{mnp}$  the $56$ gauginos and ${\bar C}_{mnpq}$ the $70$ real scalars. These fields are local in the coordinates  
\bea
y~=~\,(\,x,\,{\bar x},\,{x^+},\,y^-_{}\equiv {x^-}-\,\frac{i}{\sqrt 2}\,{\theta_{}^m}\,{{\bar \theta}^{}_m}\,)\ .
\eea

\ndt  The superfield $\phi$ and its complex conjugate $\bar\phi$ satisfy
\be
d^m\,\phi\,(\,y\,)\,=\,0\;\; ;\qquad {\bar d}_n\,{\bar \phi}\,(\,y\,)\,=\,0\ ,\quad \,{\phi}\,=\,\frac{1}{4}\,\frac{{(d\,)}^8}{{\parp}^4}\,{\bar \phi}\ ,
\ee
\noindent where
\bea
d^{\,m}\,=\,-\,\frac{\partial}{\partial\,{\bar \theta}_m}\,-\,\frac{i}{\sqrt 2}\,\theta^m\,\parp\;\; ;\quad{\bar d}_n\,=\,\frac{\partial}{\partial\,\theta^n}\,+\,\frac{i}{\sqrt 2}\,{\bar \theta}_n\,\parp\ ,\quad {(d\,)}^8\,\equiv\,d^1\,d^2\,\ldots\,d^8\ .
\eea

\vskip 0.5cm

\noindent The kinematical, spectrum generating, supersymmetry generators are~\cite{Bengtsson:1983pg}, 
\bea
q^m_{\,+}\,=\,-\,\frac{\partial}{\partial\,{\bar \theta}_m}\,+\,\frac{i}{\sqrt 2}\,\theta^m\,\parp ;\qquad {\bar q}_{\,+\,n}=\;\;\;\frac{\partial}{\partial\,\theta^n}\,-\,\frac{i}{\sqrt 2}\,{\bar \theta}_n\,\parp\ ,
\eea
satisfying $\{\ q^m_+\ , \bar q_{+n}\ \}\ =\ i\sqrt 2 \delta^m_n\ \parp$, while the dynamical ones are given by
\be
\label{dynsus}
q^{\,m}_-\,=\, \frac{\bar \del}{\parp} \,q^m_{\,+}\,  ,\quad
{\bar q}_{-n}\,=\,\frac{\partial}{\parp}\,{\bar q}_{\,+\,n}\, .
\ee
These satisfy the free $\mathcal N=8$ supersymmetry algebra closing to the Hamiltonian generator
\bea
\{\ q^m_-\ , \bar q_{-n}\ \}\ =\ i\sqrt 2 \delta^m_n\ \frac{\partial \bar \partial}{\parp}\ .
\eea

\ndt
In the interacting theory, the dynamical generators pick up corrections order by order thus generating the interacting Hamiltonian.
\vskip 0.5cm

\subsection*{The action to order $\kappa$}
\vskip 0.1cm

\noindent To order $\kappa$, the action for ${\mathcal N}=8$ supergravity reads~\cite{Ananth:2005vg}

\be
\label{n=8}
\beta\,\int\;d^4x\,\int d^8\theta\,d^8 \bar \theta\,{\cal L}\ ,
\ee
where $\beta\,=\,-\,\frac{1}{64}$ and
\bea
\label{one}
{\cal L}&=&-\bar\phi\,\frac{\Box}{\partial^{+4}}\,\phi\ + \frac{4}{3}\ \kappa \left( \frac{1}{\delp^4} \bar \phi\ {\bar \partial} \bar \del \phi\ \delp^2 \phi\ -\ \frac{1}{\delp^4} \bar \phi\ \delp  \bar \del \phi\ \delp \bar \del \phi \ +\ c.c.\right) .
\eea
\vskip 0.2cm
\noindent The d'Alembertian is
\bea
\label{dal}
\Box\,=\,2\,(\,\partial\,{\bar \partial}\,-\,\partial_+\,\parm\,)\ ,
\eea
\noindent $\kappa\,=\,{\sqrt {8\,\pi\,G}}$ and Grassmann integration is normalized such that $\int d^8\theta\,{(\theta)}^8=1$.
\vskip 0.5cm
\ndt The correction to the dynamical supersymmetry generator at this order is

\be\label{Q}
\bar q_{-m}{}^{(\kappa)} \phi=  \frac{1}{\parp}(\bar \partial \bar q_{m} \phi {\parp}^2 \phi - \parp \bar q_{m} \phi \parp \bar \partial \phi)\ ,
\ee
where the $+$ sign on the kinematic supersymmetery generators is no longer shown.

\vskip 0.5cm

\subsection{$E_{7(7)}$ symmetry }
\vskip 0.1cm

\ndt The non-linear $\E/SU(8)$  transformations to order $\kappa$ are given by \cite{Brink:2008qc}
\begin{align}\label{E}
\delta \phi=&~-\frac{2}{\kappa}\,\theta^{klmn}_{}\,\overline\Xi^{}_{klmn}\nn \\
&+\frac{\kappa}{4!}\,\Xi^{mnpq}_{}\frac{1}{\partial^{+2}}\left(\overline d_{mnpq} \frac{1}{\partial^+}\phi\,\partial^{+3}_{}\phi \, -\,4\,\overline d_{mnp} \phi\,\overline d_q\partial^{+2}_{}\phi \,+\, 3\,\overline d_{mn} {\partial^+}\phi\,\overline d_{pq}\partial^{+}_{}\phi \right),
\end{align}
where $\theta^{klmn}=\theta^k\theta^l \theta^m\theta^n$, $\overline d_{m_1...m_n} = \bar d_{m_1}....\bar d_{m_n}$ and $\overline\Xi^{}_{klmn} = \frac{1}{2} \epsilon_{klmnpqrs} \,\Xi^{pqrs}$, a constant. These $70$ coset transformations along with the linear $SU(8)$ transformations
\be
\label{SU8}
T^m{}_n ~=~ \frac{i}{2\sqrt{2} \,\d^+} \left( q^m \bar q_n\, -\,\frac{1}{8}\,\delta^m{}_n\, q^p \bar q_p \right)\  ;\quad  [\,T^m{}_n\,,\,T^p{}_q\,] ~=~\delta^p{}_n\, T^m{}_q - \delta^m{}_q \,T^p{}_n \ ,
\ee

\ndt
 constitute the entire $E_{7(7)}$ algebra. In compact coherent state-like notation the transformation (\ref{E}) can be written
\begin{equation}
\delta \phi~=~
-\frac{2}{\kappa}\,\theta^{mnpq}_{}\,\overline\Xi^{}_{mnpq}\,+\,
\frac{\kappa}{4!}\,\Xi^{mnpq}  \left(\frac{\d}{\d\eta}\right)_{mnpq}\frac{1}{\partial^{+2}}\left(e^{\eta \hat{\bar d}} \partial^{+3} \phi\, e^{-\eta \hat{\bar d}}\partial^{+3} \phi \right)\Bigg|_{\eta=0} + \m O(\kappa^2),
\end{equation}
where 

$$
\eta\hat{\bar d} = \eta^m\frac{\bar d_m}{\partial^+},~~{\rm and}~~\left(\frac{\d}{\d\eta}\right)_{mnpq} \equiv~ \frac{\d}{\d\eta^m}\frac{\d}{\d\eta^n}\frac{\d}{\d\eta^p}\frac{\d}{\d\eta^q}\ .
$$
\vskip 0.2cm
\ndt
This formulation is particularly useful for checking the commutation relations with other generators like the superPoincar\'e ones.
Note that in this formalism, the $E_{7(7)}$ symmetry, which is a duality symmetry of the vector fields and a non-linear $\sigma$-model symmetry of the scalar fields in the covariant formalism, transforms all the physical fields in the supermultiplet. Hence the supermultiplet is a representation of both the superPoincar\'e algebra as well as of the $E_{7(7)}$ one leading us to question which is the more basic one.

\vskip 0.5cm
\section{Maximal supergravity in $d=3$ - version I}
\label{withcubic}
\vskip 0.1cm
\ndt {\it {obtained by dimensional reduction from $(\mathcal N=8, d=4)$ supergravity}}
\vskip 0.5cm
\ndt When we dimensionally reduce the $d=4$ theory to $d=3$, we are left with the dependence on one transverse derivative, $ \del$. We obtain, for the action for the $d=3$ theory (up to an overall constant)

\be
\mathcal S\ =\ \int d^3 x\  d^8 \theta\  d^8 \bar \theta\ \mathcal L\ ,
\ee
\ndt 
where
\be \label{d=3 L}
\mathcal L\ =\ - \bar{\phi}\ \frac{\Box}{\delp^4}\ \phi \ + \ \frac{4}{3}\ \kappa \left( \frac{1}{\delp^4} \bar \phi\ {\partial}^2 \phi\ \delp^2 \phi\ -\ \frac{1}{\delp^4} \bar \phi\ \delp  \del \phi\ \delp  \del \phi \ +\ c.c.\right)\ ,
\ee
\vskip 0.2cm
\ndt where the $\Box$ here, is the three-dimensional d'Alembertian (see also appendix A). Before we study the symmetries of this action, we divert our attention to the $E_{8(8)}$ invariant supergravity theory in $d=3$. This theory does not admit vertices of odd order ($\kappa$, $\kappa^3$ etc.), due to the $SO(16)$ $R$-symmetry. The action of the linear $SO(16)$ and its non-linearly realised quotient $E_{8(8)}/SO(16)$ on the light-cone superfield $\phi$ was extensively studied in~\cite{Brink:2008hv}.

\vskip 0.5cm
\section{Maximal supergravity in $d=3$ - version II}
\label{withoutcubic}
\vskip 0.1cm
\ndt {\it {the manifestly $E_8$-invariant version}}
\vskip 0.3cm
\ndt There is a better known form for maximal supergravity in three dimensions. We discuss this version in this section, before relating it to the form in Section \ref {withcubic}. Maximal supergravity in three dimensions is invariant under an $E_{8(8)}$ symmetry. The same chiral superfield $\phi$ introduced earlier describes all the degrees of freedom: 128 bosons and 128 fermions,
\be
{\bf 256}\ =\ {\bf 128_b}\ +\ {\bf 128_f}
\ee
\ndt The action for this theory contains no three-point coupling, since three spinor representations cannot form a scalar.
\vskip 0.2cm
\ndt The linear action of $\bar q_m$, $q^m$ on the superfield 
\be
\delta_{\bar s}^{kin} \phi(y)\ =\ \bar \epsilon_m q^m \phi(y)\ ,\quad \delta_s^{kin} \phi(y)\ =\ \epsilon^m \bar q_m \phi(y)
\ee
yield the kinematical light-cone supersymmetries, with $\epsilon$ being the parameter.
\vskip 0.5cm

\subsection*{$SO(16)$ invariance of the theory}
\vskip 0.2cm
\ndt In $\mathcal N=8$ superspace, the Grassmann variables, $\theta^m$ and $\bar \theta_m$, form a ${\bf 16}$ representation
$$
SO(16)~\supset~ SU(8)\,\times\, U(1)\ ,\qquad {\bf 16} ~=~ \bf8\,+\,\overline{\bf 8}\ .
$$
 The quadratic action of the $q^m$, $\bar q_m$ generaors on $\phi$ generates the 120 $SO(16)$ transformations, which are decomposed in terms of $SU(8) \times U(1)$ as follows.
\be
{\bf 120}\ =\ {\bf 63_0}\ +\ {\bf 28_{-1}}\ +\ \overline{{\bf 28}}_{\bf 1}\ +\ {\bf 1_0}
\ee
The $SU(8)$ generators are given in ({\ref{SU8}}) and $U(1)$ generators are given by~\cite{Brink:2008hv}

\begin{equation}\label{U1}
 T~=~\frac{i}{4 \sqrt{2} \, \d^+}\,[\, q^{m}_{}\,,\,\bar q_{m}\, ]\ ,\ \quad \ [\,T\,,\,T^m{}_n\,]~=~0\ .
\ee

\ndt
The coset transformations $SO(16)/(SU(8)\times U(1))$ are generated by the ${\bf 28}$ and ${\bf \overline{28}}$ of $SU(8)$ 

\begin{equation}\label{28}
T^{mn}~=~\frac{1}{2}\frac{1}{\d^+} q^mq^n\, , \qquad T_{mn}~=~\frac{1}{2}\frac{1}{\d^+} \bar q_m \bar q_n\ ,
\end{equation}
which close on ($SU(8) \times U(1)$) 
\begin{equation*}
\ [\, T^{mn}\,,\, T_{pq}\,  ]~ = ~\delta^n{}_p T^m{}_q \,-\,
\delta^m{}_p T^n{}_q\, -\,\delta^m{}_q T^n{}_p  \,+\,\delta^m{}_q T^n{}_p \, +\, 2\,(\,\delta^n{}_p\delta^m{}_q \,-\,\delta^n{}_q \delta^m{}_p \,)\, T \ .
\end{equation*}
Hence, the linear $SO(16)$ transformations read

\[
 \delta^{}_{SU_8} \, \varphi~=~ \omega^n{}_{m}\,T^{m}{}_{n}\, \varphi\ ,~\quad \delta_{U(1)}\, \varphi ~=~ T \,\varphi\ ,
 \]
\begin{equation} \label{lin SO(16)}
\delta_{\bf 28}\,\varphi~=~ \alpha_{mn}\,\frac{q^mq^n}{\d^+}\, \varphi\ , \qquad 
\delta_{\bf \overline{28}}\,\varphi ~=~ \alpha^{mn} \frac{\bar q_m \bar q_n}{\d^+}\, \varphi \ ,
\end{equation}
where $\omega^n{}_{m}$, $\alpha_{mn}$, and $\alpha^{mn}$  are the transformation parameters.

\vskip 0.5cm

\subsection{$E_{8(8)}$ symmetry}
\vskip 0.2cm
We decompose the non-linearly realized coset $E_{8(8)}/SO(16)$  in terms of $SU(8)\times U(1)$ representations
\be
\bf128 = \bf1_{2}'\,+\, \bf28'_{1} \,+\,\bf70_{0}\,+\,\overline{\bf28}'_{-1} \,+\,\bf\bar1'_{-2}\;\;\ .
\ee
We identify the $\bf 70$ as the representation in  $\E/SU(8)$; the rest of the coset $E_{8(8)}/SO(16)$ transformations form two $U(1)$ singlets, a twenty-eight dimensional representation ${\bf 28'_1}$ and its complex conjugate $\overline{{\bf 28'}}_{\bf -1}$, (which are not related to the ${\bf 28}$ and $\overline{\bf 28}$ of the $SO(16)$ discussed previously). All the bosonic components of the superfield contain a constant term in the $E_{8(8)}/SO(16)$ variation, just as in a $\sigma$-model.
\vskip 0.2cm

\ndt All the 128 $E_{8(8)}/SO(16)$ coset transformations can be expressed in a compact form~\cite{Brink:2008hv}
\begin{align}\nonumber \label{E8 coset}
&\delta^{}_{E_{8(8)}/SO(16)}\,\phi~=~\frac{1}{\kappa}\,F\,+\,\kappa\,\epsilon^{m_1m_2 \dots m_8}\,\sum_{c=-2}^{2}
\left(\hat{\overline d}_{m_1m_2\cdots m_{2(c+2)}} \partial^{+c}_{}\,F\right)\\
&\quad\times
\Bigg\{ \left(\frac{\delta}{\delta\, \eta} \right)_{m_{2c+5}\cdots m_8}\,\partial^{+(c-2)} \left( e^{\eta\cdot \hat{\bar d} }  \,\partial^{+(3-c)}\phi\, e^{-\eta\cdot \hat{\bar d} } \partial^{+(3-c)}\phi\,\right)\bigg|_{\eta=0}
\,+\, \m O(\kappa^2)\Bigg\},
\end{align}
where the sum is over the $U(1)$ charges $c=2,1,0-1,-2$ of the bosonic fields, and 

\begin{eqnarray}
F&=&\,\frac{1}{{\partial^+}^2}\,\beta\,(y^-)\,\,+\,i\,\theta^{mn}_{}\,\frac{1}{\d^+}\,{\overline \beta}_{mn}\,(y^-)-\,\theta^{mnpq}_{}\,{\overline \beta}^{}_{mnpq}\,(y^-)+\nn \\
&&+\,i\widetilde\theta^{}_{~mn}\,\d^+\,\beta^{mn}\,(y^-)+\,{4}\,\widetilde\theta\,{\d^+}^2\,{\bar \beta}\,(y^-)\ ,\nn
\end{eqnarray}
and
$$\hat{\overline d}_{m_1m_2\cdots m_{2(c+2)}} ~\equiv~ \hat{\overline d}_{m_1}\hat{\overline d}_{m_2}\cdots\hat{\overline d}_{2(c+2)}\ .$$ 
It is remarkable that the $E_{8(8)}$ symmetry can be represented on the same supermultiplet as the $\E$ symmetry.
\vskip 0.5cm

\section{Relating the two different versions of three-dimensional maximal supergravity}

\ndt Having described the two different forms of maximal supergravity in three dimensions, we are now in a position to establish a link between them. We will relate the $d=3$ action with a three-point coupling (\ref{d=3 L}), obtained from dimensionally reducing $(\mathcal N=8, d=4)$ supergravity to the $E_{8(8)}$ invariant supergravity theory sans a three-point coupling. We will do this through a field redefinition and show that the dimensionally reduced form is also invariant under $SO(16)$ transformations, which are now non-linearly realized on the superfield.
\vskip 0.5cm

\subsection{The field redefinition}
\vskip 0.1cm

\ndt
The Lagrangian for the $SO(16)$ invariant theory reads
\be \label{3D L}
\mathcal L\ =\ -\ \bar \phi\ \frac{\Box}{\delp^4}\ \phi\ +\ \mathcal O(\kappa^2) \ .
\ee 
We want a field redefintion that will map the kinetic term in (\ref{3D L}) to a kinetic term plus the $\it O(\kappa)$ terms in (\ref{d=3 L}). Based on dimensional analysis, we start with the ansatz
\be 
\phi\ =\ \phi' + \alpha \ \kappa\ \delp^A\ ( \delp^B \phi'\ \delp^C \phi' )\ +\ \beta\ \kappa \ \delp^D\ (\delp^E \phi' \delp^F \bar \phi') \ ,
\ee
\vskip 0.2cm
\ndt where $\alpha$, $\beta$ are constants to be determined and the integers A, B, C, D, E, F obey
\be
A+B+C = 2\ ,\quad D+E+F = 2\ .
\ee 
Simple computations lead us to
\be
\label{redef}
\phi \rightarrow \phi\ =\ \phi'\ +\ \frac{1}{3}\ \kappa\ (\delp \phi'\ \delp \phi')\ +\ \frac{2}{3} \kappa\ \delp^4 \left( \frac{1}{\delp^3}\ \phi'\ \delp \bar \phi' \right)\ ,
\ee
which correctly reproduces the cubic terms in (\ref{d=3 L}) as shown in appendix A. The $(\phi' \bar \phi')$ piece in the field redefinition achieves the same effect as replacing $\del^-$ by $\frac{\del^2}{\delp}$ in the interaction terms. We thus arrive at the new Lagrangian
\be \label{new L}
\mathcal L'\ =\ - \bar{\phi'}\ \frac{\Box}{\delp^4}\ \phi' \ + \ \frac{4}{3}\ \kappa \left( \frac{1}{\delp^4} \bar \phi'\ {\partial}^2 \phi'\ \delp^2 \phi'\ -\ \frac{1}{\delp^4} \bar \phi'\ \delp  \del \phi'\ \delp  \del \phi' \ +\ c.c.\right)\ ,
\ee
\vskip 0.2cm
\ndt
which exactly matches (\ref{d=3 L}), since $\phi' = \phi$ at lowest order. Thus the the dimensionally reduced action for $d=3$ maximal supergravity with a cubic vertex can be obtained from the $SO(16)$-invariant action (without a cubic vertex) by a field redefinition.

\vskip 0.5cm
\subsection{$SO(16)$ symmetry revisited}
\vskip 0.2cm
\ndt
The linear action of the various $SO(16)$ generators on $\phi$ is listed in (\ref{lin SO(16)}).  The SO(16) invariance of the Lagrangian (\ref{3D L}) at the free order implies
\be \label{C}
\delta \mathcal L\ = \ -\ (\delta \bar \phi)\ \frac{\Box}{\delp^4}\ \phi\ -\ \bar \phi\ \frac{\Box}{\delp^4}\ (\delta\phi)\ =\ 0 
\ee
\ndt
(Note: $\delta_{SO(16)} \phi\,\equiv\,\delta \phi$ for simplicity.)
\vskip 0.3cm
\ndt
To understand the action of $SO(16)$ on the new superfield $\phi'$, we invert (\ref{redef}) to obtain

\be \label{A1}
\phi'\ =\ \phi\ -\ \frac{1}{3}\ \kappa\ (\delp \phi\ \delp \phi)\ -\ \frac{2}{3}\ \kappa\ \delp^4\ \left( \frac{1}{\delp^3}\phi\ \delp \bar \phi \right)\ ,
\ee
\be
\delta \phi' \ =\ \delta \phi\ -\ \frac{2}{3}\ \kappa\ ( \delp (\delta \phi)\ \delp \phi)\  -\ \frac{2}{3}\ \kappa\ \delp^4 \left( \frac{1}{\delp^3}(\delta \phi)\ \delp \bar \phi \right) -\ \frac{2}{3}\ \kappa\ \delp^4 \left( \frac{1}{\delp^3}\phi\ \delp (\delta \bar \phi) \right)\ . \nn
\ee
\vskip 0.2cm
\ndt In appendix B, we prove that the new Lagrangian in (\ref {new L}) is also $SO(16)$ invariant albeit in a non-linear fashion. Finally, in appendix C, we prove that this new theory is also $E_8$ invariant. 
\vskip 0.3cm
\ndt {\it {Thus  the $d=3$ supergravity Lagrangian with cubic interaction vertices, obtained by dimensional reduction from $(\mathcal N=8, d=4)$ supergravity, is equivalent to the $d=3$ Lagrangian without cubic vertices and futher, both these versions have an $E_{8(8)}$ symmetry.}} 
\vskip 0.3cm

\section{Oxidation back to $d=4$ preserving the  $E_{8(8)}$ symmetry}
\vskip 0.2cm
\ndt  We now demonstrate how the $d=3$ Lagrangian, without cubic vertices, may be oxidized to four dimensions while preserving the $E_{8(8)}$ symmetry. We achieve this by introducing a ``new" tranverse derivative, $\del_2$. 
\vskip 0.3cm
\ndt In~\cite{Brink:2008hv}, the $E_{8(8)}$ symmetry was used  to construct the order-$\kappa^2$ dynamical supersymmetry transformations in $d=3$
\bea
&\delta_s^{dyn}\,\phi& =\, \epsilon^m \frac{\del}{\delp}\ \bar q_m\,\phi\ \nn \\
&& +\  \frac{\kappa^2}{2} \sum^2_{c=-2}\ \frac{1}{{\delp}^{(c+4)}} \Bigg\{ \frac{\delta}{\delta a}\ \frac{\delta}{\delta b} \left( \frac{\delta}{\delta \eta}\right)_{m_1 m_2 ...m_{2(c+2)}} \bigg( E \delp^{(c+5)} \phi\  E^{-1} \bigg) \Bigg|_{a=b=\eta=0}\nn \\
&& \times \frac{\epsilon^{m_1 m_2 ...m_8}}{(4-2c)!} \left( \frac{\delta}{\delta \eta} \right)_{m_{2c+5}...m_8}\ \delp^{2c}\ \bigg( E \delp^{(4-c)} \phi E^{-1}\ \delp^{(4-c)} \phi \ \bigg) \Bigg|_{\eta=0} \Bigg\}\ ,\nn \\
&&
\eea
\ndt
where 
\be
E \equiv e^{a \hat{\del}\,+\,b\epsilon \hat{\bar q}\,+\,\eta \hat{\bar d}}\ \ \text{and}\ \ E^{-1} \equiv  e^{-a \hat{\del}\,-\,b\epsilon \hat{\bar q}\,-\,\eta \hat{\bar d}}\ , \nn
\ee
\ndt
with
\be
a\,\hat{\del}\,=\, a\frac{\del}{\delp}\ ,\quad b\,\epsilon \hat{\bar q}\,=\, b\, \epsilon^m\frac{\bar q_m}{\delp} ,\quad \eta \hat{\bar d}\,=\, \eta^m \frac{\bar d_m}{\delp} \ . \nn
\ee
\vskip 0.2cm
\ndt We oxidize this expression to $d=4$ by replacing all the $\partial\ ( \ =\partial_1)$ by a generalized derivative 
\be
\nabla\ \equiv\ \partial_1\ +\ i\ \partial_2\ ,
\ee
\ndt such that
\be
\left[\delta_s^{dyn}\,\phi\,(\del, \delp, \bar q_m, \bar d_m, \phi)\right]_{d=3}\ \longrightarrow\; \ \left[\delta_s^{dyn}\,\phi (\nabla, \delp, \bar q_m, \bar d_m, \phi)\right]_{d=4} \, . \nn
\ee
\vskip 0.3cm
\ndt
We now note that for maximally supersymmetric theories one can obtain the light-cone Hamiltonian using the quadratic form expression~\cite{Ananth:2006fh}

\be
\mathcal H\ =\ \frac{1}{4\sqrt 2}\,(\mathcal{W}_m, \mathcal{W}_m) \,\equiv\, \frac{2i}{4 \sqrt 2}\,\int d^8 \theta\ d^8 \bar \theta\ d^3 x\ \overline{\mathcal W}^m\ \frac{1}{{\partial^+}^3}\ \mathcal W_m\ ,
\ee

\ndt
where $\mathcal{W}_m$ is the dynamical supersymmetry variation on $\phi$

\be
\delta_s^{dyn}\,\phi\, \equiv\, \epsilon^m\, \mathcal W_m\, .
\ee

\ndt
Once we obtain $\mathcal W_m$ in $d=4$ through the oxidation, we can in principle construct a $SO(16)$ invariant Hamiltonian with only even order coupling. In doing so, we need to take the complex conjugate of $\mathcal W_m$, which will introduce the conjugate ``new" derivative
\be
\partial_1\ -\ i\ \partial_2\ \equiv \overline{\nabla}\ .
\ee
\vskip 0.3cm
\ndt
This method of oxidation respects both the $SO(16)$ and the full $E_{8(8)}$ symmetry, because the generalized derivatives, $\nabla$ and $\overline \nabla$ do not contain any $q^m\, , \bar q_m$ or $d^m\, , \bar d_m$ operators, which can affect the invariance of the Hamiltonian in $d=4$. Thus, we will end up with a maximal supergravity theory in $d=4$ with the same field content as in (\ref {superfield}). Since the $\mathcal N=8$ theory is unique, we have arrived at a form of $\mathcal N=8$ supergravity in $d=4$ with $E_{8(8)}$ symmetry to this order. One could ask why we had to leave four dimensions in the first place? We could have simply found a field redefinition from the $\mathcal N=8$ theory to a form that is $E_{8(8)}$-invariant. The answer is that our procedure, of going down one dimension, allows us to render the enhanced symmetry manifest. This manifest enhanced symmetry is the difficult step to achieve. Once this is in place, we oxidize the theory, preserving the enhanced symmetry arriving at our goal.
\vskip 0.3cm
\ndt Note that in order to argue that the Hamiltonian is $E_{8(8)}$-invariant to this order we must treat the states as $128$-dimensional spinors. These are clearly not the four-dimensional states of $(\mathcal N=8, d=4)$ supergravity. In order to argue that this symmetry is present in $d=4$ scattering amplitudes we must add up such amplitudes such that the external states span the full $128$-dimensional spinors. We have seen though as in the original paper on complete one-loop amplitudes~\cite{Green:1982sw} that all of those amplitudes have the same divergence pattern. It is a further assumption that this is true also to higher loops which the analysis in~\cite{Bern:2011qn} indicates.
\vskip 0.3cm
\ndt We have not discussed the supersymmetry generators in this form of  $(\mathcal N=8, d=4)$ supergravity. Since this formulation is obtained by a field redefinition from the original one we do not expect supersymmetry generators to be straightforward to write down. This is a price we have to pay in this formalism which is minimal in terms of field components.

\section{Conclusions}

\ndt Maximally supersymmetric gravity and Yang-Mills theories have been found to have the simplest perturbation series among theories of their kind. In some sense they have only the bare bone structure needed to build a perturbation series which is both unitary and causal. There are also strong reasons to believe that the perturbation series of $(\mathcal N=8, d=4)$ supergravity is, in a certain sense, the square of that in $(\mathcal N=4, d=4)$ Yang-Mills theory (KLT-relations~\cite{Kawai:1985xq,AT}).This perturbative simplicity is all the more remarkable since at least the Yang-Mills theory non-perturbatively even knows about superstring theory through the AdS-CFT duality.
\vskip 0.3cm
\ndt Our analysis does not shed light on whether the $(\mathcal N=8, d=4)$ supergravity is perturbatively finite. We can only argue that the perturbation series ought be more finite than what the usual counterterm arguments based on $E_{7(7)}$ and maximal supersymmetry suggest (for a related discussion, see~\cite{BN}). Counterterms could in principle be constructed in our formalism, but this is a formidable task that we hope to return to. Not only do we need to construct counterterms, we must also prove that they cannot be absorbed by a field redefinition. Further, the remaining counterterms should be invariant under the residual reparametrization, local supersymmetry and gauge symmetries as we have shown in the case of pure gravity~\cite{Bengtsson:2012dw}. We can only see two ways to finally settle the question (of finiteness): do the full calculation or find a power counting argument as was achieved in the case of $\mathcal N=4$ Yang-Mills theory~\cite{Brink:1982wv} and its deformations~\cite{AKS1,AKS2}. That analysis cannot be carried over straightforwardly but new additional symmetry-related inputs may help limit the possible diagrams that need checking.
\vskip 0.3cm
\ndt Are there even larger symmetries lurking in these theories? There have been strong indications that the affine algebras $E_{10}$ and $E_{11}$ could be present~\cite{Damour:2002cu,West:2001as}. Such symmetries could possibly be realized by the superfield and all its superspace derivatives. However, we find it difficult to see how such symmetries could directly be symmetries of the scattering amplitudes since that would amount to infinitely many kinematical constraints on the amplitudes. Those symmetries must be more deeply ingrained in these theories and we hope to return to this question in future publications.

\vskip 1cm
\ndt {\it \bf {Acknowledgments}}
\vskip 0.1cm

\ndt The work of SA is partially supported by a DST-SERB grant (EMR/2014/000687). SM acknowedges support from a CSIR NET fellowship.

\vskip 1cm
\newpage

\appendix
\section{Verification of field redefinition}

\vskip 0.2cm
\ndt
Under the field redefinition (\ref{redef}), the kinetic term in (\ref{d=3 L}) becomes

\begin{eqnarray}
-\ \bar \phi\ \frac{\Box}{\delp^4} \phi &=& -\ 2\ \bar \phi\ \frac{(\del^2\ -\ \delp \del^-)}{\delp^4} \ \phi \ \nn \\
&=&-\ 2\  \left\{ \bar \phi' \ +\ \frac{1}{3}\ \kappa\ ( \delp \bar \phi'\ \delp \bar\phi')\ +\ \frac{2}{3}\ \kappa\ \delp^4 \left( \frac{1}{\delp^3} \bar \phi'\ \delp \phi' \right) \right\} \ \times \nn \\
&& \frac{(\del^2\ -\ \delp \del^-)}{\delp^4}\ \left\{  \phi'\ +\ \frac{1}{3}\ \kappa\ (\delp \phi'\ \delp \phi')\ +\ \frac{2}{3} \kappa\ \delp^4 \left( \frac{1}{\delp^3}\ \phi'\ \delp \bar \phi' \right) \right\} \nn
\end{eqnarray}

\ndt
The free order term gives back the kinetic term. Now, at order $\kappa$ we keep terms which are of the form $\bar \phi' \phi' \phi'$ only. \footnote{The other kind of terms  $\bar \phi' \bar \phi' \phi'$, which are just complex conjugate of these terms, reproduce the  $\kappa\ \bar \phi' \bar \phi' \phi'$ vertex in (\ref{d=3 L})}

\be
\label{new terms}
- \ \frac{2}{3}\ \kappa\ \bar \phi'\  \frac{(\del^2\ -\ \delp \del^-)}{\delp^4} \ (\delp \phi'\ \delp \phi')\ -\ \frac{4}{3}\ \kappa\ \delp^4 \left( \frac{1}{\delp^3} \bar \phi'\ \delp \phi' \right)\frac{(\del^2\ -\ \delp \del^-)}{\delp^4}\ \phi'\ =\ \mathcal{A}\ +\ \mathcal{B} \nn
\ee
\vskip 0.2cm
\ndt
$\mathcal{A}$ and $\mathcal{B}$ can be further simplified as follows.
\bea
\mathcal{A}&=&- \ \frac{2}{3}\ \kappa\ \frac{1}{\delp^4}\ \bar \phi'\ (\del^2\ -\ \delp \del^-) \ (\delp \phi'\ \delp \phi') \nn \\
&=& -\ \frac{4}{3}\ \kappa\ \frac{1}{\delp^4}\ \phi'\ (\delp \del^2 \phi'\ \delp \phi'\ +\ \delp \del \phi'\ \delp \del \phi')\  +\ \frac{4}{3}\ \kappa\ \frac{1}{\delp^4}\ \bar \phi' \ \delp (\delp \del^- \phi'\ \delp \phi') \nn \\
&& \nn \\
\mathcal{B}&=& -\ \frac{4}{3}\ \kappa\ \left( \frac{1}{\delp^3} \bar \phi'\ \delp \phi' \right) (\del^2\ -\ \delp \del^-)\ \phi' \nn \\
&=& +\ \frac{4}{3}\ \kappa\ \frac{1}{\delp^4}\ \bar \phi'\  \delp\ (\del^2 \phi'\ \delp \phi')\ -\ \frac{4}{3}\ \kappa\ \frac{1}{\delp^4}\ \bar \phi' \ \delp\ (\delp \del^- \phi'\ \delp \phi') \nn \\
&=& +\ \frac{4}{3}\ \kappa\ \frac{1}{\delp^4}\ \bar \phi'\  (\delp \del^2 \phi'\ \delp \phi'\ +\ \del^2 \phi'\ \delp^2 \phi')\ - \frac{4}{3}\ \kappa\ \frac{1}{\delp^4}\ \bar \phi' \ \delp\ (\delp \del^- \phi'\ \delp \phi') \nn \\
&&\nn
\eea
Hence, the order-$\kappa$ terms are
\be 
\mathcal{A}\ +\ \mathcal{B}\ =\  \frac{4}{3}\ \kappa \left( \frac{1}{\delp^4} \bar \phi'\ {\partial}^2 \phi'\ \delp^2 \phi'\ -\ \frac{1}{\delp^4} \bar \phi'\ \delp  \del \phi'\ \delp  \del \phi' \ \right) \nn ,
\ee
\vskip 0.2cm

\newpage

\section{$SO(16)$-invariance of the new Lagrangian}

The $SO(16)$ variation of $\mathcal L'$ yields

\be
\delta \mathcal L'\ =\ \delta {\mathcal L'}_{kinetic}\ +\ \delta {\mathcal L'}_{cubic} \ ,
\ee

\ndt
where

\be
\delta \mathcal L'_{kinetic}\ = \ -\ (\delta \bar \phi')\ \frac{\Box}{\delp^4}\ \phi'\ -\ \bar \phi'\ \frac{\Box}{\delp^4}\ (\delta\phi')\ 
\ee

\ndt and

\bea
 \label{F}
\delta {\mathcal L'}_{cubic}&=&  + \ \frac{4}{3}\ \kappa \Bigg( \frac{1}{\delp^4} \bar (\delta \phi')\ {\partial}^2 \phi'\ \delp^2 \phi'\ +   \frac{1}{\delp^4} \bar \phi'\ {\partial}^2 (\delta \phi')\ \delp^2 \phi'\ +\ \frac{1}{\delp^4} \bar \phi'\  \del^2 \phi' \delp^2 (\delta \phi')\nn \\
&&-\ \frac{1}{\delp^4} (  \delta \bar \phi')\ \delp  \del \phi'\ \delp  \del \phi' -\ 2 \ \frac{1}{\delp^4} \bar \phi'\ \delp  \del(\delta \phi')\ \delp  \del \phi' \Bigg)\ +\ c.c. \ .
\eea
\vskip 0.3cm
\ndt
Using (\ref{A1}) and keeping terms up to order $\kappa$, we get

\bea
\label{B1}
\delta \mathcal L'_{kinetic} &=&  \left\{-\ (\delta \bar \phi)\ \frac{\Box}{\delp^4}\ \phi\ -\ \bar \phi\ \frac{\Box}{\delp^4}\ (\delta\phi) \right\}\ \nn
\\
&&+ \Bigg\{ \frac{1}{3} \ \kappa\ (\delta \bar \phi)\ \frac{\Box}{\delp^4}\ (\delp \phi\ \delp \phi)\ +\ \frac{2}{3}\ \kappa\   \bar \phi\ \frac{\Box}{\delp^4}\ (\delp (\delta \phi)\ \delp \phi)\ \nn \\
&& +\ \frac{2}{3}\ \kappa\ \delp^4 \left( \frac{1}{\delp^3} (\delta \bar \phi) \ \delp \phi \right) \frac{\Box}{\delp^4}\phi\  + \frac{2}{3}\ \kappa\ \delp^4 \left( \frac{1}{\delp^3} \bar \phi \ \delp (\delta \phi) \right) \frac{\Box}{\delp^4}\phi\   \nn \\
&&+ \frac{2}{3}\ \kappa\ \delp^4 \left( \frac{1}{\delp^3} \bar \phi \ \delp \phi \right) \frac{\Box}{\delp^4}(\delta \phi )\Bigg\} +\ c.c 
\eea
\ndt
The terms of order $\kappa^0$ cancel against each other, as in eq. (\ref{C}). We have only considered terms of the form $(\bar \phi \phi \phi)$, since the others are contained in the complex conjugate. After partially integrations of $\delp$ and simple manipulations, (\ref{B1}) takes the form
\bea
\delta \mathcal L'_{kinetic}&=& -\ \frac{4}{3}\ \kappa \Bigg( \frac{1}{\delp^4} \bar (\delta \phi')\ {\partial}^2 \phi'\ \delp^2 \phi'\ +   \frac{1}{\delp^4} \bar \phi'\ {\partial}^2 (\delta \phi')\ \delp^2 \phi'\ +\ \frac{1}{\delp^4} \bar \phi'\  \del^2 \phi' \delp^2 (\delta \phi')\nn \\
&&-\ \frac{1}{\delp^4} (  \delta \bar \phi')\ \delp  \del \phi'\ \delp  \del \phi' -\ 2 \ \frac{1}{\delp^4} \bar \phi'\ \delp  \del(\delta \phi')\ \delp  \del \phi' \Bigg)\ +\ c.c. \ ,
\eea
\ndt

\vskip 0.3cm
\ndt

\ndt
which cancels against (\ref{F}) rendering the new Lagrangian, with a cubic vertex, SO(16)-invariant. 

\newpage

\section{$E_8$ invariance}

\ndt We show here, how the non-linearly realised $SO(16)$ for the action with a three-point coupling can be extended to an $E_{8(8)}$ symmetry. The action of the 128 $E_{8(8)}/SO(16)$ transformations on the superfield $\phi$ is given in (\ref{E8 coset}). We know that two such coset transformations should close on $SO(16)$, (we denote the coset transformations here by $\delta' \phi$)

\be
[\delta'_1, \delta'_2] \phi\ =\ \delta_{SO(16)}\ \phi\ .
\ee
\vskip 0.2cm
\ndt
 Using (\ref{A1}), we can readily express $\delta' \phi'$  in terms of $\delta' \phi$. Let us consider two coset transformations, $\delta'_1$ and $\delta'_2$ on $\phi'$

\bea
[\delta'_1, \delta'_2]\ \phi' \ &=&\ [\delta'_1, \delta'_2]\ \phi\ -\ \frac{1}{3}\ \kappa\ [\delta'_1, \delta'_2] ( \delp \phi\ \delp \phi)\ \- \frac{2}{3}\ \kappa\ [\delta'_1, \delta'_2] \left\{{\delp}^4 \left( \frac{1}{{\delp}^3}\phi\ {\delp} \bar \phi \right)\right\} \nn \\
&=& \delta_{SO(16)} \phi\ +\ \mathcal{X}\ +\ \mathcal{Y}
\eea

\ndt
where $\mathcal{X}$ and $\mathcal Y$ simplify to

\bea
\mathcal{X} &=& -\ \frac{1}{3}\ \kappa\ [\delta'_1, \delta'_2] ( \delp \phi\ \delp \phi)\ \nn\\
&=& -\frac{2}{3}\ \kappa\ [\delp (\delta'_1 \delta'_2 \phi) \delp \phi\ +\ \delp(\delta'_2 \phi) \delp(\delta'_1 \phi)\ -\ \delp(\delta'_2\delta'_1 \phi) \delp \phi\ -\ \delp(\delta'_1 \phi) \delp(\delta'_2 \phi)] \nn \\
&=& -\frac{2}{3}\ \kappa\ (\delp [\delta'_1, \delta'_2] \phi\ \delp \phi) \nn \\
& =& -\frac{1}{3}\ \kappa\ \delta_{SO(16)} (\delp \phi\ \delp \phi) \nn
\eea

\ndt
and
\be
\mathcal{Y}\ =\ - \frac{2}{3}\ \kappa\ [\delta'_1, \delta'_2] \left\{{\delp}^4 \left( \frac{1}{{\delp}^3}\phi\ {\delp} \bar \phi \right)\right\}\ =\ -\frac{2}{3}\ \kappa\ \delta_{SO(16)}  \left\{{\delp}^4 \left( \frac{1}{{\delp}^3}\phi\ {\delp} \bar \phi \right)\right\} . \nn
\ee
\vskip 0.3cm
\ndt
So, from (\ref{A1}) we find that for the new field $\phi'$
\bea
[\delta'_1, \delta'_2] \ \phi'\ &=& \ \delta_{SO(16)} \phi\ - \frac{1}{3}\ \kappa\ \delta_{SO(16)} (\delp \phi\ \delp \phi) -\  \frac{2}{3}\ \kappa\ \delta_{SO(16)}  \left\{{\delp}^4 \left( \frac{1}{{\delp}^3}\phi\ {\delp} \bar \phi \right)\right\} \nn \\
&=& \delta_{SO(16)} \phi'\ ,
\eea
thus proving that the transformations close.

\end{document}